# EEG-based Reaction Time prediction with Fuzzy Common Spatial Patterns and Phase Cohesion using Deep Autoencoder based data fusion

Vivek Singh
Department of Electronics and Communication Engineering
Indian Institute of Technology,
Roorkee, India
vivek_s@ece.iitr.ac.in

Tharun Kumar Reddy
Department of Electronics and Communication Engineering
Indian Institute of Technology,
Roorkee, India
tharun.reddy@ece.iitr.ac.in

**Abstract**— Drowsiness state of a driver is a topic of extensive discussion due to its significant role in causing traffic accidents. This research presents a novel approach that combines Fuzzy Common Spatial Patterns (CSP) optimised Phase Cohesive Sequence (PCS) representations and fuzzy CSP-optimized signal amplitude representations. The research aims to examine alterations in Electroencephalogram (EEG) synchronisation between a state of alertness and drowsiness, forecast drivers' reaction times by analysing EEG data, and subsequently identify the presence of drowsiness. The study's findings indicate that this approach successfully distinguishes between alert and drowsy mental states. By employing a Deep Autoencoder-based data fusion technique and a regression model such as Support Vector Regression (SVR) or Least Absolute Shrinkage and Selection Operator (LASSO), the proposed method outperforms using individual feature sets in combination with a regressor model. This superiority is measured by evaluating the Root Mean Squared Error (RMSE), Mean Absolute Percentage Error (MAPE), and Correlation Coefficient (CC). In other words, the fusion of autoencoder-based amplitude EEG power features and PCS features, when used in regression, outperforms using either of these features alone in a regressor model. Specifically, the proposed data fusion method achieves a 14.36% reduction in RMSE, a 25.12% reduction in MAPE, and a 10.12% increase in CC compared to the baseline model using only individual amplitude EEG power features and regression.

**Index Terms**— Data fusion, Phase Cohesive Sequence (PCS), Reaction Time (RT), Autoencoder, Root Mean Squared Error (RMSE), Mean Absolute Percentage Error (MAPE) and Correlation Coefficient (CC) INTRODUCTION

## I. INTRODUCTION

The word" drowsiness" (or" fatigue") describes the propensity to feel sleepy or nod off. According to a WHO report published in 2013 [1], drowsy driving accounts for almost 6% of all traffic accidents worldwide. This statistic has been referred to as a critical reason why drowsiness detection systems in vehicles have become so popular. Due to its strong correlation with both mental and physical activities, EEG is regarded as one of the most reliable indicators of physiological signals. In the literature [2]–[6], a number of techniques for using EEG to detect fatigue have been put forth. These techniques can be broadly divided into two categories: amplitude-based and phase-based. We begin by summarising the amplitude-based methods. In their analysis of four algorithms, Budi et al. [7] found that drowsiness led to an elevation in the rate of total spectral power in theta with respect to alpha frequency bands compared to beta band power. The Level of Session Generalizability (LSG) is a concept that Wei et al. [8] introduced in order to enhance the performance of drowsiness detection using a Transfer Learning (TL) based method. Other authors [8]– [10] suggested using type-2 fuzzy classifiers and the motor coordination phase to detect cognitive impairment in driving, as well as regression with Random Forest and integrated fatigue metrics according to power spectrum density along with sample entropy analysis and using motor planning phase and type-2 fuzzy classifiers to detect the cognitive failure in driving.

The second class of approaches rely on phase information to detect drowsiness. Phase Cohesion (PC) or phase-based analysis is discussed as a fundamental concept for neuronal information processing and communication within and between different brain regions. Various methods for measuring phase synchrony, such as the Phase Locking Value (PLV) [11], Momentary Phase Difference (MPD) [12], and Mean Phase Coherence (MPC) [13], are described. This paper proposes a new approach called PCS, which integrates Temporal Phase difference with fuzzy CSP for fatigue detection. Although extensive research focuses on EEG classification problems, there has been a notable lack of attention given to EEG regression problems. Only a limited number of studies have been conducted on tasks such as continuous workload level estimation and reaction time prediction.

The proposed method is unique in the following ways:

- The use of phase-based features for EEG-based drowsiness detection problems is rarely explored in the existing literature. The proposed method addresses this gap by utilising PCS feature representations to train intelligent models.

- In the literature, autoencoders are employed for unsupervised deep learning. In this work, for the first time, we realise deep data fusion using autoencoders.
- Extensive testing of the proposed data fused model on the individual subjects and on the combination of subjects.

The paper is organised as follows: Section II is dedicated to the proposed method. Section III evaluates a novel Phase Cohesive Sequence Representation to predict Reaction Time. Section IV provides the Results and Discussion details. Concluding remarks are provided in section V.

## II. PROPOSED METHOD

Let D denote the number of channels and K denote the number of time samples per trial, where $X_m \in R^{D \times K}$, $m \in \{1,2,..,M\}$ represents the m th EEG trial. Trial $X_m$ is a band pass filtered signal where the mean has been excluded from each channel. In order to adapt the approach to regression machine learning problem, we came up with fuzzy classes using the idea of fuzzy sets (assume R fuzzy classes). Then, R + 1 regions are created using the interval range [0,100], and the boundary section points are denoted by {$y_{pr}$ }, r = {1,.., R} respectively.

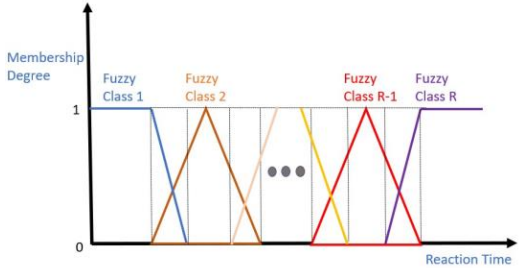

Fig. 1:' R' fuzzy classes to train RT values with triangular fuzzy membership

$$P_r = \frac{100r}{R+1}, r = 1, \dots, R \quad (1)$$

Every $y_{pr}$ is the Pr percentile for the training data of Reaction Times. The next step is to introduce R fuzzy classes and then the training of all RT values for each of the R fuzzy classes, similar to how M classes would be used in a multi-class classification scenario.

A particular $y_m$ would belong to each fuzzy class at a membership degree $\in [0,1]$. Also, for every fuzzy class, we calculate an average covariance matrix as follows:

$$\bar{\Sigma}_r = \frac{\sum_{m=1}^{M} \mu_R(y_m) X_m X_m^T}{\sum_{m=1}^{M} \mu_R(y_m)}, m = 1, \dots, M \quad (2)$$

Where $\mu_r(y_m)$ represents the $y_m$'s membership degree of $r^{th}$ fuzzy class.

The One Vs Rest (OVR) CSP is developed to adapt the CSP from binary categorisation to R classes. CSPr discovers a matrix $A^*_r \in R^{D \times H}$ for class r, where H is the number of spatial filters, to maximise class r's variance compared to the rest.

$$A_R^* = argmax \frac{Tr(A^\top \bar{\Sigma}_r A)}{Tr\left(A^T \sum_{j \neq m} \bar{\Sigma}_j A\right)} \quad (3)$$

The mean covariance matrix for trials of class r is denoted as $\bar{\Sigma}r$ . The concatenation of the H eigenvectors and the H greatest eigenvalues of the matrix $\left(\sum_{i \neq m} \bar{\Sigma}_i\right)^{-1} \bar{\Sigma}_r$ is known as $A^*_r$. We combine the H filters that were obtained for any of the R classes to produce $A^* = [A^*_1, \cdots, A^*_R] \in R^{D \times RH}$. A spatially translated trial can then be calculated using the formula $X_m' = A^{*T} A_m$, m = 1,..., M.

### A. Phase Synchronisation Index (PSI)

PSI serves as a statistical tool for exploring the synchronisation of the brain's electrical activity in EEG data, providing a momentary interconnectivity index. Cognitive tasks arise because of the collaboration among different functional areas located in varied segments of the brain. The coupling observed connecting these areas during task performance is the synchronisation of the brain's electrical activity. PSI and its derived variations are widely used as measures of phase synchronisation. Lets assume, the temporal phases of two signals, $x_1(t)$ and $x_2(t)$, are $\theta_1(t)$ and $\theta_2(t)$. According to [17], the one trial PSI (sPSI) of a particular trial may be described as follows:

$$sPSI = \left| \frac{1}{M_s} \sum_{t=1}^{M_s} e^{|\theta_1(t) - \theta_2(t)|} \right| \quad (4)$$

where Ms denotes the trial's sample quantity in numbers. The analytical signal obtained from the Hilbert transform can be used to determine the instantaneous phase θ(t). For every random signal w(t), the analytic signal y(t) is represented as:

$$y(t) = w(t) + i\widetilde{w}(t) \quad (5)$$

$$\widetilde{w}(t) = \frac{1}{\pi} \int_{-\infty}^{\infty} \frac{w(\tau)}{t - \tau} d\tau \quad (6)$$

Where the Hilbert transform of w(t) is $\widetilde{w}(t)$. The momentary phase θ(t) can be obtained by:

$$\theta(t) = arctan\left(\frac{\widehat{w}(t)}{w(t)}\right) \quad (7)$$

The sPLV has a value between 0 and 1, with the extreme values corresponding to unsynchronised and fully synchronised signals, respectively.

### B. Phase Cohesive Sequence representations

The temporal Phase Difference sequence (TPD), Δθ(t) between two different signals $s_1(t)$ and $s_2(t)$ is expressed as:

$$\Delta\theta(t) = |\theta_1(t) - \theta_2(t)| \quad (8)$$

The concept of variation of instantaneous phasors and sPSI were combined by the authors in [12]. They also developed a framework for estimating a linear transform that maximises temporal phasor variance within one class while concurrently minimising it within the another. Like CSP, this proposed way explicitly leverages phase information for binary categorisation. A novel framework known as Fuzzy CSP for regression (FuzzyCSPr) was developed with the goal of enhancing the variance of temporal phase differences within each fuzzy class while concurrently reducing the phase discrepancy across the other fuzzy class. This framework incorporates an analogy to the regression algorithm [15]. The discovered features, known as PCS representations, are derived from the TPD sequence, which is employed to evaluate the synchronisation among EEG signals.

*Optimisation of Fuzzy Spatial Filters:* Multi-class CSP is extended to challenges with regression using fuzzy sets in Fuzzy CSPr [15]. As previously stated, due to spatial smearing and blurring effects, the SNR of the EEG signals from individual channels is frequently low. Hence, it appears wise to calculate spatial filters that maximise the variance of momentary phase within a specific fuzzy category while minimising it across the other fuzzy classes. This approach aims to generate more distinctive PCS characteristics for the EEG classification of fuzzy classes. Therefore, we suggest optimising the spatial filters to maximise the final PCS features discriminative power. It can be represented as:

$$A_i^* = argmax \frac{(A^\top \bar{\Sigma}_{\Delta\theta_i} A)}{\left(A^T \sum_{j \neq i} \bar{\Sigma}_{\Delta\theta_j} A\right)} \quad (9)$$

where TPD sequence mean covariance matrices for fuzzy classes i and j are represented by $\bar{\Sigma}_{\Delta\theta_i}$ and $\bar{\Sigma}_{\Delta\theta_j}$, respectively. The spatial cohesion filters are column vectors of matrix A. The TPD sequence of an EEG trial is used to extract features within a specific frequency range using a method akin to the fuzzy CSP algorithm. The features are referred to as' PCS' representations.

## III. EVALUATING PHASE COHESIVE SEQUENCE REPRESENTATIONS TO PREDICT REACTION TIME

Based on EEG data for a lane-control task, the proposed feature representations are evaluated and compared to alternative phase-based standard approaches for predicting reaction times (RT).

### A. Lane-Control Task

EEG data obtained from 30 electrode locations that were positioned using a modified version of the internationally recognised 10-20 electrode placement system. The Institutional Review Board of Veterans General Hospital authorised the study in Taipei, Taiwan. To examine the EEG patterns associated with alterations in attention and performance during real-world drowsiness scenarios, a group of 12 university students (with an average age of 22.4 and a standard deviation of 1.6) from National Chiao Tung University (NCTU) in Taiwan willingly participated in data collection over five months. [16]. Driving simulations were conducted using virtual reality (VR) technology on a dynamic driving simulator. The driving simulations consisted of both motion sessions and motionless sessions. During the motion sessions, a car frame was affixed to a 6 Degree of Freedom (DoF) Stewart motion platform synchronised with the driving scene. Conversely, the motion platform remained idle and inactive during the motionless sessions. The VR driving scene simulated nighttime driving at 120 km/h on a straight, unoccupied highway with two lanes in each direction. The computer program generated random disruptions, known as deviation onsets, causing the car to drift towards either side of the cruising lane with the same probability.

**TABLE I: No. of Trials (Numtrials) and Average RT in the dataset**

| Subjects | 1 | 2 | 3 | 4 | 5 | 6 |
|---|---|---|---|---|---|---|
| Numtrials | 628 | 568 | 653 | 350 | 362 | 361 |
| Average RT | 0.897 | 1.668 | 0.925 | 0.881 | 1.070 | 1.145 |
| Subjects | 7 | 8 | 9 | 10 | 11 | 12 |
| Numtrials | 239 | 484 | 550 | 641 | 190 | 143 |
| Average RT | 2.020 | 1.541 | 1.001 | 0.704 | 1.076 | 2.677 |

Participants were instructed to promptly steer the vehicle back to the cruising lane (response onset) and, once it reached the approximate center of the lane, maintain control of the steering wheel (response offset) after each deviation. The lane change trial consisted of three distinct events: deviation onset, response onset, and response offset. Afterward, a lane-departure trial was randomly initiated around 5 to 10 seconds after the response offset of the previous trial. The subject's reaction time (RT) for each lane change trial was measured as the time between the start of the deviation and the start of the response. If a subject did not respond within 2.5 (1.5) seconds, the vehicle would collide with the left (right) side of the road as it approached the barrier.

Using a 5-second EEG trial immediately preceding the RT, the objective is to predict the reaction time.

### B. EEG Data Pre-processing

- To increase the SNR (Signal to Noise Ratio), raw EEG data was first run through EEGLAB's typical pre-processing pipeline (PREP), which primarily have 3 operations [17].

- Removal of line noise.
- Identifying and eliminating substantial reference signals.
- Interpolation of unreliable channels.

- Subsequently, the raw data were downsampled to a frequency of 250 Hz.

- Following the downsampling step, the raw data was segmented into 5-second trials. Specifically, if the lane change event occurs at the time' t', the [t − 5,t] EEG data is used for forecasting the reaction time (RT). Each EEG trial has a size of 30×1250.

- To remove outliers in the reaction time (RT) values, EEG trials with RT values surpassing the addition of the averages and three times the standard deviation are excluded.

- The collected trials are subsequently filtered using a finite impulse response (FIR) band-pass filter with a frequency range of 1 to 20 Hz.

- Finally, the processed data is applied to suitable spatial filters.

*C. Reaction Time Pre-processing*

Similar to section IV D of of the paper [15], the RT values of twelve subjects were pre-processed. Subject 12's data collection was inaccurate and contained anomalies in the data recording, so it was excluded from further analysis. This is due to the fact that many response times exceeded 5 seconds, which is highly absurd in actual usage. After pre-processing, the complete distribution of all RTs is displayed in Fig.2.

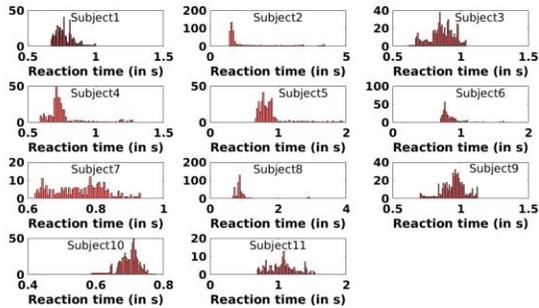

Fig. 2: Distribution of RT values

*D. Feature Evaluation*

For every probable fusion of the features and regression method, the regression performance is calculated using 8-fold cross-validation. For every EEG trial, the corresponding feature sets are determined.

1. Utilising band-pass filtered EEG trials, theta and alpha powerband features are extracted using band-pass filtered EEG trials. The average power spectral density within the theta band (4-8 Hz) and the alpha band (8-13 Hz) for each channel was computed using Welch's method. These $30 \times 2 = 60$ band powers were converted to decibels (dBs) and used as our features, denoted as 'FS1'.

2. The band-pass filtered EEG trials were utilised to extract proposed PCS features. We employed twentyone spatial filters (H = 21) for each fuzzy class and three fuzzy sets (K = 3) for each of the reaction times (RTs). The resulting feature vector, PCS, had a size of (63×1) = 63. These features are referred to as 'FS2'.

Here, 'FS1 is termed to be an Amplitude-based feature vector (because the power is extracted from signal amplitude), while 'FS2' is termed a Phase-based feature vector. To determine the final reaction time value, a pair of the feature sets obtained in the prior phase is fused via a deep autoencoder and passed on to a fully connected layer with a regression node.

*E. Performance Metrics*

The metrics used to determine the regression performance are RMSE, CC, and MAPE. Let's say there are N training points, ydi represents the i th data point's true RT value and yi indicates the predicted RT value.

## IV. RESULTS AND DISCUSSION

The EEG lane-keeping task's reaction time dataset is used to conduct the experiments. The optimal choices for the no. of fuzzy classes and the no. of filters for every fuzzy class are obtained as K=3 and F=21, respectively. The preprocessed EEG RT dataset is used to extract the PCS-FCSP features.

To demonstrate the deep autoencoder's superior data fusion abilities, we compared with and without the autoencoder after passing the fused/unfused feature pairs ('FS1', 'FS2') through traditional regression models such as LASSO Regularized Regression and Ridge regression models. We implemented LASSO and Ridge regression models in MATLAB R2016 and used the sci-kit-learn SVR (Support vector regression) tool respectively to run LASSO and SVR simulations. Our simulations showed that SVR is inefficient in handling large-scale training datasets due to its algorithmic complexity and storage capacity requirements. Furthermore, when confronted with high-dimensional input features, SVR's generalisation ability is limited. LASSO and SVR were optimised through an inner 8-fold cross-validation on the training dataset. In our experiments, the proposed approaches utilising deep data fusion demonstrated superior performance compared to baseline individual feature methods regarding RMSE, CC, and MAPE. We employed a deep autoencoder-based data fusion model, which was trained for 110 epochs using an Adagrad optimiser with a learning rate of 0.14. Table I and II provides the average results of these experiments.

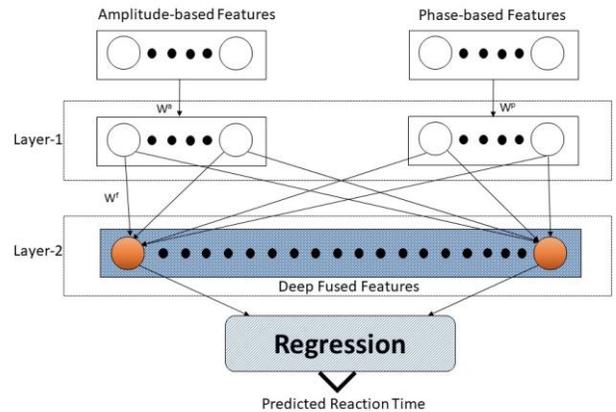

Fig. 3: Amplitude and Phase-based features fusion using Deep Autoencoder

TABLE II: Comparison of regression metric for deep autoencoder-based data fusion over standard regression models

| Model | AE-LASSO | $FS1-LASSO$ | $FS2-LASSO$ |
|---|---|---|---|
| RMSE | **0.0504** | 0.0630 | 0.0675 |
| CC | **0.820** | 0.810 | 0.731 |
| MAPE | **9.50** | 10.37 | 17.03 |
| Model | AE-SVR | $FS1-SVR$ | $FS2-SVR$ |
| RMSE | **0.0720** | 0.0603 | 0.0592 |
| CC | **0.635** | 0.653 | 0.749 |
| MAPE | **16.28** | 11.51 | 10.84 |

TABLE III: Two-way ANOVA results deep autoencoder-based data fusion versus others

| | RMSE | CC | MAPE |
|---|---|---|---|
| $p-value$ | 0.014 | 0.000 | 0.003 |

The primary contribution of this study is introducing a novel drowsiness detection system that utilises deep data fusion of Phase Cohesive Sequence (PCS) and amplitude representations for regression. This approach is integrated into the fuzzy-CSP framework (PCS-FCSP) and validated using an EEG-based lane-keeping task dataset, employing 8-fold cross-validation. The statistical analysis and regression outcomes provide evidence of the effectiveness of the proposed PCS-FCSP representations and the system. As future work, the authors intend to integrate PCS features with Regularized Fuzzy-CSP using the Joint Approximate Diagonalization (JAD) framework. This study applies data fusion learning to mitigate the risk of overfitting and enhance performance.

## V. CONCLUSION

The paper proposes a novel deep data-fusion framework based on PCS features (PCS features are extracted from TPD sequences) and Amplitude features to predict reaction time (RT) in an EEG-based lane-keeping task. The proposed framework outperformed powerband features with better RMSE, correlation coefficient (CC), and MAPE values when passed through LASSO and SVR after Deep-data fusion. The proposed Data fusion with deep autoencoder approach achieved the best performance with 14.36% smaller RMSE, 25.12% smaller MAPE, and 10.12% larger CC than SVR. The study proposes that PCS-data fusion is beneficial in BCI applications, especially for online adaptation with limited training data and as an initial step towards developing large-scale Multi-task learning BCIs. Additionally, we highlight the potential exploration of deep data-fusion semi-supervised multi-task approaches in future research.

## VI. ACKNOWLEDGEMENT

This research work is funded by the Science and Engineering Research Board (SERB), Department of Science & Technology, Government of India, under project number SER-1968-ECD and sanction order number SRG/2022/001886.